\newcommand{\pl}{\partial}
\newcommand{\be}{\begin{equation}}\newcommand{\ee}{\end{equation}}
\newcommand{\bea}{\begin{eqnarray}}\newcommand{\eea}{\end{eqnarray}}
\newcommand{\nn}{\nonumber}\newcommand{\p}[1]{(\ref{#1})}

\documentstyle[12pt]{article}
\begin{document}
\thispagestyle{empty}
\begin{flushright}ENSLAPP-L-415-92\\
December 1992
 \end{flushright}
\vskip 1.0truecm
\begin{center}{\bf\Large
$N=4$ super KdV equation}\end{center}
 \vskip 1.0truecm
\centerline{\bf F. Delduc${}^{(a)}$ and E. Ivanov${}^{(a,b)}$}
\vskip 1.0truecm
\centerline{${}^{(a)}$ \it Lab. de Phys. Th\'eor. ENSLAPP, ENS Lyon}
\centerline{ 46 All\'ee d'Italie, 69364 Lyon, France}
\vskip5mm
\centerline{${}^{(b)}$\it Laboratory of Theoretical Physics, JINR,}
\centerline{Dubna, Head Post Office, P.O. Box 79, 101000 Moscow, Russia}
\vskip 1.0truecm  \nopagebreak

\begin{abstract}
We construct $N=4$ supersymmetric KdV equation as a hamiltonian flow
on the $N=4\;SU(2)$ super Virasoro algebra. The $N=4$ KdV superfield,
the hamiltonian and the related Poisson structure are concisely formulated
in $1D \;N=4$ harmonic superspace. The most general hamiltonian is
shown to necessarily involve $SU(2)$ breaking parameters which are
combined in a traceless rank 2 $SU(2)$ tensor. First nontrivial conserved
charges of $N=4$ super KdV (of dimensions 2 and 4) are found to
exist if and only if the $SU(2)$ breaking tensor is
a bilinear of some $SU(2)$ vector with a fixed length proportional to
the inverse of the central charge of $N=4\;SU(2)$ algebra. After the
reduction to $N=2$ this restricted version of $N=4$ super KdV
goes over to the $a=4$ integrable
case of $N=2$ super KdV and so is expected to be integrable.
We show that it is bi-hamiltonian like its $N=2$ prototype.

\end{abstract}
\newpage\setcounter{page}1

{\bf 1. Introduction.} The Korteweg-de Vries (KdV) hierarchy, being
the text-book example of $2D$
integrable system, now proved to be intimately related to many areas of
modern theoretical physics, including conformal field theories,
$2D$ gravity, matrix models, etc. As was observed in \cite{{a1},{a2}},
it admits
a nice interpretation  as a hamiltonian flow on the Virasoro algebra.
In this approach the KdV field is identified with
a $2D$ conformal stress-tensor and the KdV equation, as well as next
equations from the KdV hierachy, come out as
hamiltonian equations with respect to the Poisson brackets forming a
classical Virasoro algebra.

In [3-7] integrable $N=1$ and $N=2$ super KdV equations
have been
deduced as hamiltonian flows on the $N=1$ and $N=2$ super Virasoro
(superconformal) algebras (SCA in what follows). An intriguing peculiarity
of the $N=2$ case is that there exists a one-parameter family of
the super KdV equations associated with $N=2$ SCA, but
only for three selected values of the parameter, $a=-2, 4, 1$, these
equations are integrable, i.e. possess an infinite number of the
conserved quantities in involution and a Lax representation
[6-8].

A natural minimal extension of $N=2$ SCA is the $N=4 \;SU(2)$ SCA
\cite{a9}. Like the former it is generated by the
currents with canonical spins $1, 3/2, 2$, the spinor currents forming
a complex doublet of the internal symmetry group $SU(2)$.
This is to be contrasted, e.g., with the $N=3$ SCA, the odd sector of which
involves a real $SO(3)$ triplet of the spin $3/2$ currents and a
singlet spin $1/2$ current \cite{a9}. In the context of conformal field theory
the $N=4\;SU(2)$ SCA was treated, e.g., in \cite{{et},{a13}}. An obvious
interesting
problem
is to construct the $N=4$ super KdV associated with the $N=4\;SU(2)$
SCA and to see whether the three integrable $N=2$ super KdV
equations
can be promoted to integrable $N=4$ ones \footnote{Super KdV equations
associated
with the $N=3$ and $N=4\;SO(4)$ super Virasoro algebras were discussed in
ref. \cite{{luk},{bik},{yung}}.}.
This could shed more light
on the origin of integrability in super KdV hierarchies and on the
meaning of the parameter $a$.

In the present letter we solve the first part of this problem and report on
some partial results regarding  the
second one. We construct, in a manifestly covariant $N=4$ superfield form,
the most general supersymmetric KdV equation for which $N=4\;SU(2)$ SCA
provides the second hamiltonian structure and argue, by examining the
question of existence of higher order conserved charges, that at least one
integrable $N=2$ super KdV equation, with $a=4$, generalizes to
$N=4$ while preserving integrability.
The parameter $a$ reveals an unexpected
meaning within the $N=4$ framework: the $N=4$ super KdV equation necessarily
contains $SU(2)$ breaking parameters which are combined into a traceless rank
2 $SU(2)$ tensor and $a$ is recognized as a component of this tensor. The
requirement of the existence of non-trivial conserved
charges in involution beyond the hamiltonian (we explicitly construct
the dimension 2 and dimension 4 charges) restricts the components of the
$SU(2)$ breaking tensor to be
bilinear in
components of some vector having a fixed length and so
parametrizing a sphere $S^2 \sim SU(2)/U(1)$. The length (radius of the
sphere) turns out to be proportional to the inverse of the central charge of
 the $N=4\;SU(2)$
SCA. After performing the reduction $N=4 \Rightarrow N=2$ one ends up
with the integrable $N=2$ super KdV equation corresponding to $a=4$. All the
conserved
charges of the latter (an infinite tower of them) seem to have proper $N=4$
counterparts thus implying the $N=4$ super KdV with the aforementioned
restrictions on the parameters to be
integrable. One more
argument in favour of its integrability
is based on its bi-hamiltonian nature: we present the first hamiltonian
structure for it.

Throughout the paper we systematically use the $1D \;N=4$ version of the
harmonic superspace approach \cite{a10}. This allows us to formulate
all our results in a concise form and to avoid some technical difficulties
inevitable when working in ordinary $1D \;N=4$ superspace.
\vspace{0.5cm}

{\bf 2. N=4 SU(2) superconformal algebra in harmonic superspace.} In ordinary
$1D \;N=4$ superspace with the coordinates $Z^M \equiv
(x, \theta^i,\bar\theta_j)$,
where $i,j=1,2$ are $SU(2)$ doublet indices, the currents generating
$N=4\;SU(2)$ SCA are accomodated into
a dimension 1 superfield $V^{ij}( Z^M )$,
$V^{ij}=V^{ji}$, satisfying the constraints (see, e.g., ref. \cite{a11}):
\begin{equation}
D^{(i}V^{jk)}=0\;,\ \ \bar D^{(i}V^{jk)}=0\;.
\label{ct1}
\end{equation}
Here
\be
D_i = \frac{\partial}{\partial \theta^i} - \frac{i}{2} \bar{\theta}_i
\frac{\partial}{\partial x}\;, \; \bar{D}^i = - \frac{\partial}{\partial
\bar{\theta}_i}
+ \frac{i}{2} \theta^i \frac{\partial}{\partial x} \;, \;\;
\{ D_i, \bar{D}^j \} = i \delta^i_j \partial \;, \; \{ D_i, D_j \}
=0 \label{ct1a}
\ee
and the $SU(2)$ indices are raised and lowered by the antisymmetric tensors
$\epsilon^{ij}\;, \epsilon_{ij}$ ($\epsilon^{ij}\epsilon_{jk} = \delta^i_k\;,
\epsilon_{12} = -\epsilon^{12} = 1$). It is straightforward to check that
the constraints \p{ct1} leave in $V^{ij}$ only the following independent
superfield projections
\be
w^{ij} = i\;V^{ij}\;, \;\; \xi^{k} = D^i V^k_i\;,\;\; T(x) = {i\over 3}
\bar{D}^iD^k V_{ik}\;, \label{ct1b}
\ee
where the numerical factors are inserted for further convenience.
The $\theta$ independent parts of these projections,
$w^{ij} (x),\;\xi^{l}(x),\; T(x)$, up to unessential rescalings coincide
with the currents of $N=4\;SU(2)$ SCA: the $SU(2)$ triplet of the spin 1
currents generating $SU(2)$ Kac-Moody subalgebra, a complex doublet of the
spin 3/2 currents and the spin 2 conformal stress-tensor, respectively.

Let us see how the same supercurrent is represented in the harmonic
$1D\;N=4$ superspace, an extension of $\{ Z^M \}$ by the harmonic variables
$u^\pm_i$ describing a 2-sphere $\sim SU(2)/U(1)$
$$
\{ Z^M \} \Rightarrow \{ Z^M\;, \;u^{+i}\;, \;u^{-j} \}\;,
$$
\be
u^{+i}u_{i}^{-} = 1\;, \;\;\; u^{+}_iu^-_j - u^-_iu^+_j = \epsilon_{ij}
\label{ct1c}
\ee
(see \cite{{a10},{dis}} for the basics of the harmonic superspace approach).

In what follows we will need
the derivatives in harmonic variables which are given by
\be D^{++}=u^{+i}{\pl\over\pl u^{-i}}\;,\  D^{--}=u^{-i}{\pl\over\pl u^{+i}}
\;,\ D^0=[ D^{++}, D^{--}] = u^{+i}{\pl\over\pl u^{+i}} -
u^{-i}{\pl\over\pl u^{-i}}\;. \label{ct1d} \ee
The role of $D^{0}$ is to count the $U(1)$ charge of functions on the
harmonic superspace defined as the difference between the numbers of the
indices $+$ and $-$. The strict preservation of this $U(1)$ charge is one of
the basic postulates of the harmonic superspace approach. It expresses the
fact that
the harmonic variables belong to the sphere $S^2$ and the harmonic
superfields are functions on this sphere as well. Also,
instead of $D^i\;,\;\bar{D}^j$ we will use their projections
on $u^{\pm i}$
\be
D^\pm=D^iu^\pm_i\;, \;\;
\bar D^\pm=\bar D^iu^\pm_i\;, \nn \\
\ee
nonvanishing (anti)commutators of which with themselves and the harmonic
derivatives $D^{++}$, $D^{--}$ are
\bea
\{ D^- ,\bar D^+ \} &=& i\pl \;,\  \{ D^+ ,\bar D^- \}\;=\;-i\pl \;,
\label{ct1e} \\
\left[ D^{++}, D^- \right] &=& D^+\;, \;\; \left[ D^{--}, D^{+} \right]\;=\;
D^- \;. \label{ct1f}
\eea

Define now the $1D\;N=4$ harmonic superfield $V^{++}(Z,u)$ subject to the
constraints
\bea
D^+V^{++} &=& 0\;, \;\;\;\;\bar  D^+V^{++}\;=\;0 \label{ct2} \\
D^{++} V^{++} &=& 0\;. \label{ct3}
\eea
It follows from the harmonic constraint \p{ct3} that $V^{++}$
is a homogeneous function of degree 2 in $u^{+i}$
\be
V^{++} (Z,u) = V^{ij}(Z)\; u^+_iu^+_j \;. \label{ct3a}
\ee
Then, in view of the arbitrariness of $u^{+i}, \;u^{+j}$, the constraints
\p{ct2}
imply for $V^{ij}$ the original constraints \p{ct1}. Thus the
superfield $V^{++}$ obeying \p{ct2}, \p{ct3} represents the $N=4\;SU(2)$
conformal supercurrent in the harmonic $1D\;N=4$ superspace \footnote{The
harmonic
superspace form of the $N=4\;SU(2)$ conformal supercurrent has
been earlier given in ref. \cite{a12}.}.

The constraints \p{ct2} can be viewed as Grassmann analyticity conditions
covariantly eliminating in $V^{++}$ the dependence on half of the original
Grassmann coordinates, namely, on their $u^-$ projections $\theta^{-} =
\theta^{i}u^{-}_i,\;\bar{\theta}^- = \bar{\theta}^iu^-_i$. So $V^{++}$ is
an {\it analytic} harmonic superfield living on an analytic subspace
containing only the $u^+$ projections of $\theta^i\;,\;\bar{\theta}^j$.
This harmonic analytic superspace is closed under the
action of $1D \;N=4$ supersymetry (and actually under the transformations
of the whole $N=4\; SU(2)$ SCA), so one may construct additional
superinvariants as integrals over this superspace.
This opportunity will be exploited in next Sections. In what follows
we will never actually need to know the explicit coordinate structure of the
analytic superspace and how $V^{++}$ is expressed there. We will only make
use of the
constraints \p{ct2}, \p{ct3} and some important consequences of them, e.g.
\be
(D^{--})^3 V^{++} = 0\;, \;\; D^-(D^{--})^2 V^{++} = \bar D^{-}(D^{--})^2
V^{++}
=0\;,\;\;etc. \label{ct3b}
\ee

After we have represented the $N=4\;SU(2)$ supercurrent as a harmonic
superfield $V^{++}$, it remains to write the Poisson bracket between two
$V^{++}$'s which yields the $N=4\;SU(2)$ SCA Poisson brackets for the
component currents. Surprisingly, it is almost uniquely determined
by dimensionality and compatibility with
the constraints \p{ct2}, \p{ct3}. It reads
\bea  \left\{ V^{++}(1),
 V^{++}(2)\right\} &=& {\cal D}^{(++|++)} \Delta (1 - 2) \nn \\
{\cal D}^{(++|++)} &\equiv & (D^+_1\bar D^+_1)(D^+_2\bar D^+_2)\left(
\left[ {1\over 2}D^{--}_1 - \left({u^-_1u^+_2
\over u^+_1u^+_2}\right)\right]
 V^{++}(1)
+{k\over 4}\pl_{1}\right),
\label{poi}\eea
where $\Delta(1-2)=\delta(x_1-x_2)\;(\theta^1-\theta^2)^4$ is the
ordinary $1D \;N=4$ superspace delta function, and we refer to \cite{dis}
for
more details on harmonic distributions \footnote{Actually, the harmonic
singularity in the r.h.s. of \p{poi} is fake: it is cancelled after
decomposing the harmonics $u^{\pm i}_2$ over $u^{\pm i}_1$ with making
use of the
completeness relation \p{ct1c}.}. Using the algebra of spinor
and harmonic derivatives and also the completeness condition \p{ct1c},
one can check that the r.h.s of \p{poi} is consistent with the
constraints \p{ct2}, \p{ct3} with respect to both sets of arguments and
is antisymmetric under the interchange $1\Leftrightarrow 2$. To be convinced
that it gives rise to the correct Poisson brackets for the component
currents, let us, e.g., deduce from (\ref{poi})
the Poisson brackets of $SU(2)$ Kac-Moody currents.
After simple algebraic manipulations we obtain for $w^a \equiv
\sigma _i^{a\;j}w_j^{\;\;i}$ the familiar relation:
\bea
\left\{ w^{a}(1), w^{b}(2)\right\}= \epsilon^{abc}w^{c}(1)\;\delta (1-2)
+ {k\over 2} \;\delta^{ab}\;\pl_{1}\delta(1-2) \;.\eea
All other currents can also be checked to satisfy the relations needed
to constitute $N=4\;SU(2)$ SCA.

Finally, we point out that it is straightforward to rewrite the Poisson
structure \p{poi} in ordinary $1D\;N=4$ superspace.
But there it looks much more complicated:
it involves intricate combinations of $SU(2)$ indices, etc.
\vspace{0.5cm}

{\bf 3. N=4 super KdV equation from N=4 SU(2) SCA.} Our aim is to deduce the
most general super KdV equation with the second
hamiltonian structure given by the $N=4\;SU(2)$ SCA in the form \p{poi}.
The only requirement we impose beforehand is rigid $1D\;N=4$ supersymmetry.
The most general dimension 3 $N=4$ supersymmetric hamiltonian one may
construct out
of $V^{++}$ consists of two pieces
\be
H=\int [dZ]\;V^{++}(D^{--})^2V^{++}-i\int [d\zeta^{-2}]\;c^{-4}(u)\;(V^{++})^3
\;.
\label{h3}
\ee
Here $[dZ]=dx[du]\;D^-\bar D^- D^+\bar D^+$ is the integration measure of
the full
harmonic superspace and $[d\zeta^{-2}]=dx_{A}[du]\;D^-\bar D^-$ is the
integration measure of the analytic subspace. The integral over harmonic
variables is defined so that $\int [du] \;1 = 1$ and integral of any
symmetrized product of harmonics is vanishing \cite{a10}. We see that, to
balance
the $U(1)$ charges, the integral over analytic subspace should inevitably
include the harmonic monomial
$c^{-4}(u)=c^{ijkl}u^-_iu^-_ju^-_ku^-_l$ which explicitly breaks $SU(2)$
symmetry. The coefficients $c^{ijkl}$  belong to the dimension 5 spinor
representation of $SU(2)$, i.e. form a symmetric traceless rank 2 tensor, and
completely breaks the
$SU(2)$ symmetry (case (A) in what follows), unless $c^{-4}$ takes the
special form $c^{-4}(u)=(a^{-2}(u))^2$,
$a^{-2}(u)=a^{ij}u^-_iu^-_j$  (case (B)). In case (B), the symmetry
breaking parameter belongs to the dimension 3 (vector) representation of
$SU(2)$, and thus has
$U(1)$ as a little group. We point out that the presence of the trilinear
term
in the hamiltonian is unavoidable if one hopes to eventually obtain
an integrable
super KdV equation (it should somehow contain
the $N=2$ super KdV family which is integrable only providing the relevant
hamiltonian contains a trilinear term). Thus, one necessary condition for the
integrability of $N=4$ super KdV we are going to derive is that
$SU(2)$ is broken, at least down to its $U(1)$ subgroup.

Using the hamiltonian \p{h3}, we construct an evolution equation:
\be  V^{++}_t = \{H, V^{++}\}\ee
which, after a bit tedious but straightforward computations may be cast into
the following form:
\bea
 V^{++}_t&=&i D^+\bar D^+\left\{ {k\over 2}D^{--}V^{++}_{xx}
-\left[V^{++}(D^{--})^2V^{++}-{1\over 2}(D^{--}V^{++})^2\right]_x\right.\cr
&& \left. -{3\over 20}k A^{-4}(V^{++})^2_x+{1\over 2}A^{-6}(V^{++})^3
\right\}\;. \label{kdv4}
\eea
Here $A^{-4}$ and $A^{-6}$ are differential operators on the 2-sphere
$\sim SU(2)/U(1)$
\bea
A^{-4}&=&\sum_{N=1}^4(-1)^{N+1}c^{2(N-2)}{1\over N!}(D^{--})^N,\cr
A^{-6}&=&{1\over 5}\sum_{N=1}^5(-1)^{N+1}c^{2(N-3)}{(6-N)
\over N!}(D^{--})^N
\eea
and we have used the notation:
\be c^{2N-4}={(4-N)!\over 4!}(D^{++})^Nc^{-4}, \ N=0\cdots 4.\ee

The equation \p{kdv4} is the sought $N=4$ super KdV equation. It is easy
to check that its r.h.s satisfies the same constraints \p{ct2}, \p{ct3} as
the l.h.s. One might bring \p{kdv4} into a more explicit form using
the algebra
(7), (8) (e.g., the first term takes then the familiar form
$-{k\over 2}\;V^{++}_{xxx}$), however, for many reasons it is convenient
to keep
the analytic subspace projector $D^+\bar D^+$ before the curly
brackets in \p{kdv4}. Let us also note that the hamiltonian
\p{h3} and eq.\p{kdv4} can be rewritten in ordinary $N=4$ superspace, but
they look there, like the Poisson bracket \p{poi}, very intricate. For
instance, the second term in \p{h3} would involve explicit $\theta$'s so
it would be uneasy to see that it is supersymmetric. Thus the harmonic
superspace seems to provide the most appropriate framework for formulating
$N=4$ super KdV equation. The last comment concerns the presence of the
$N=4\;SU(2)$ SCA central charge $k$ in \p{kdv4}. Making in \p{kdv4}
the rescalings
$t \rightarrow bt,\;V^{++} \rightarrow b^{-1}V^{++},\; c \rightarrow bc$ we
can in principle fix this parameter at any non-zero value. However, in order
to have a clear contact with the original $N=4\;SU(2)$ Poisson
structure \p{poi} we prefer to leave $N=4$ super KdV in its original form.

Before going further we present the
bosonic core
of our super KdV equation. It consists of two coupled equations for
the fields $T$ and $w^{ij}$, first of which is an extension of the
KdV equation and the second is a three-component generalization of the mKdV
equation
\bea T_t&=&-{k\over 2}\;T_{xxx}+3\;(T^2+{3\over
10}\;T\;c_{ijkl}\;w^{ij}w^{kl})_x
+(w^{ij}_{xx}w_{ij}-{3k\over 20}\;
c_{ijkl}\;w^{ij}_{xx}w^{kl})_x\cr &&-{1\over 2}(w^{ij}_x(w_{ij})_x+{3k\over 10}
\;c_{ijkl}\;w^{ij}_{x}w^{kl}_x)_x -{3i\over
5}\;c_{ijkl}\;(w^{ij}{w^{k}}_{m}w^{lm}
_x) \eea
\bea
&&w^{ij}_t=-{k\over 2}\;w^{ij}_{xxx}+2\;(T\;w^{ij}-{3k\over 20}\;T\;
{c^{ij}}_{kl}\;w^{kl})_x
+2i\;(w^{k(i}_x{w^{j)}}_k-{3k\over 40}\;{c^{(i}}_{klm}\;
(w^{j)k}w^{lm})_x)_x\cr
&&+{3ik\over 20}\;{c^{ij}}_{kl}\;(w^{km}_x{w^l}_m)_x
-{2i\over 5}\;{c^{(i}}_{klm}\;(Tw^{j)k}w^{lm})_x
+{1\over 10}\;c_{klmn}\;(w^{ij}w^{kl}w^{mn}+4w^{ik}w^{jl}w^{mn})_x\cr
&&-{1\over 30}\;{c^{(i}}_{klm}\;(w^{j)n}_x{w^k}_nw^{lm}+w^{j)k}
w^{ln}_x{w^m}_n+2{w^{j)}}_n w^{kn}_xw^{lm})\;.
\eea
\vspace{0.5cm}

{\bf 4. Reduction to N=2 super KdV.} As a first step in analyzing the
integrability properties of \p{kdv4} we
perform now a reduction to $N=2$ supersymmetry in order to see
which kind of $N=2$ super KdV arises thereupon.

Schematically, this reduction goes as follows. One passes to the
$N=2$ notation for $N=4$ spinor derivatives
\bea
D &\equiv& D_1\;,\;\; \bar D \;\equiv \;\bar D^1\;,\;\; d \;\equiv \;D_2\;,
\;\; \bar d \;\equiv \; \bar D^2\;, \nn \\
\{ D, \bar D \} &=& i\partial\;,\;\; \{ d, \bar d \} \;=\;i\partial \label{r1}
\eea
and identifies $D,\;\bar D$ with the spinor derivatives of $N=2$
supersymmetry. Respectively, $1D \;N=4$ superspace is represented as a
product of the $1D \;N=2$ superspace with coordinates $\{ x,\;\theta \equiv
\theta^1,\;\bar \theta \equiv \bar \theta_1 \}$ and the pure Grassmann
quotient $\{ \theta^2,\;\bar \theta_2 \}$. Then one rewrites \p{kdv4} in
ordinary $1D \;N=4$ superspace using the above notation (we do not know how
to perform the reduction directly in the harmonic superspace). Further, in
all
places where the covariant derivatives $d$ and $\bar d$ of $V^{ij}$ are
encountered, one expresses them through $D,\;\bar D$ using the constraints
\p{ct1}. The crucial observation is that the whole set of currents of $N=2$
SCA is contained in the $\theta^2,\;\bar \theta_2$ independent part of the
$N=4$ superfield $V^{12}$, hence the identification
\be
\Phi \equiv V^{12} | \label{r2}\;,
\ee
where $\Phi (x,\theta, \bar \theta)$ is the $N=2$ conformal supercurrent and
$|$ means restriction to the $\theta^2,\;\bar \theta_2$ independent part.
Finally, in the $N=4$ super KdV equation prepared as explained above one
puts
$$
\theta^2 = \bar \theta_2 = 0\;, \;\; V^{11} = V^{22} = 0\;.
$$

As the result of this reduction one gets the $N=2$ super KdV equation in
the form
\be
\Phi_t = -{k\over 2} \Phi_{xxx} -3i \left[ (D\bar D - \bar DD)\Phi \Phi
\right]_x -{i\over 2} (a-1) \left[ (D\bar D - \bar D D)\Phi^2 \right]_x
- {2a\over k}\Phi^2\Phi_x \label{r3}
\ee
with
\be
a\equiv {3\over 5} k \;c^{1212}\;. \label{r4}
\ee
This equation can be brought precisely to the form given in \cite{{a6},{a7}},
redefining
\be
\partial_t = i{k\over 2} \tilde{\partial}_t\;, \;\partial_x = -i
\tilde{\partial}_x\;, \;\Phi = {k\over 2}\;\tilde{\Phi}\;,
D = {1\over 2}\;(\tilde{D}_2 + i\tilde{D}_1)\;,\;\bar D = {1\over 2}\;(
\tilde{D}_2 - i\tilde{D}_1)\;. \label{r5}
\ee
The $N=4\;SU(2)$ Poisson structure \p{poi} is reduced to the $N=2$
one
\be
\left\{ \tilde{\Phi}(1), \tilde{\Phi}(2) \right\} =
{1\over 4k} \left( \tilde{D}_1\tilde{D}_2\tilde{\partial}
 - \tilde{D}_1\tilde{\Phi} \tilde{D}_1 - \tilde{D}_2\tilde{\Phi} \tilde{D}_2
+ 2\tilde{\Phi} \tilde{\partial} + 2\tilde{\partial}\tilde{\Phi} \right)
\left( d\bar{d} \Delta(1-2) \right)|\;. \label{r6}
\ee

Thus, the reduction $N=4 \Rightarrow N=2$ in our $N=4$ super KdV
equation \p{kdv4} and the $N=4$ Poisson structure \p{poi} gives precisely the
$N=2$ Poisson structure and the well-known one-parameter family of the
$N=2$ super KdV equations. A new point is that the parameter $a$ turns out
to be related via the equation \p{r4} to the component $c^{1212}$ of the
$SU(2)$ breaking tensor present in the $N=4$ super KdV equation. Thereby,
this parameter acquires a group-theoretical meaning.
\vspace{0.5cm}

{\bf 5. Conserved charges.} As was mentioned in Introduction,
$N=2$ super KdV equation is integrable
only
for $a=4,\;-2,\;1$. Then, in view of the relation \p{r4} one may expect that
the
$N=4$ super KdV equation is integrable only under certain restrictions on
the $SU(2)$ breaking tensor $c^{ijkl}$. To see, which kind of restrictions
arises, we analyze here the question of existence of non-trivial conserved
charges for \p{kdv4} which are in involution with the hamiltonian \p{h3}.

Before all, it is clear that after the reduction to $N=2$ such
charges should go over to those of the integrable $N=2$ super KdV equations.
Any such charge of dimension, say, $l$, is known to contain in the
integrand the term $\sim (\Phi)^l$ \cite{{a6},{a7}}. These terms can be
obtained only from analytic integrals of the form
\be
\sim \int [d\zeta^{-2}]\; b^{-2(l-1)}\;  (V^{++})^l\;, \label{cc1}
\ee
where
\be
b^{-2(l-1)} = b^{i_1...i_{2(l-1)}} \;u^-_{i_1} ... u^-_{i_{2(l-1)}}\;.
\label{cc2}
\ee
Then, if the corresponding charge is required to be conserved, the highest
order contribution
to the time derivative of \p{cc1} (coming from the 3-d order term in the
r.h.s.
of \p{kdv4}) should vanish in its own right. A simple analysis shows that it
is possible if and only if
\be
b^{-2(l-1)} \sim (c^{-4})^n\;, \;\;l=2n+1 ;\;\;\;\; b^{-2(l-1)}
\sim (a^{-2})^{2n-1}\;,\; c^{-4} = (a^{-2})^{2}\;, \;\;l=2n \;. \label{cc3}
\ee
In other words, the odd dimension conserved charges can exist only provided
$b^{-2(l-1)}$ is a power of $c^{-4}$ while
the necessary condition
for the existence of the even
dimension
conserved charges
is more restrictive: $c^{-4}$ should be square of some
$a^{-2} = a^{ij}u^-_i u^-_j$, where
$a^{ij}$ is a $SU(2)$ vector, i.e. we meet the situation called in Sect. 3
the case (B).

Let us now explicitly construct several first charges. Conservation of
the dimension 1 charge :
\be H_1=\int [d\zeta^{-2}]\; V^{++}\ee
imposes no condition on the parameters of the hamiltonian.

A charge with dimension 2 exists only in the case (B) and reads:
\be H_2=\int [dz^{-2}]\; a^{-2}\;(V^{++})^2 \;.\ee
Already for this charge we find that it is conserved only
under an additional non-trivial condition on the $SU(2)$ breaking
vector $a^{ij}$: the square of the latter should be proportional to the
inverse of the central charge:
\be s\equiv a^{+2}a^{-2}-(a^0)^2={1\over 2}\;a^{ij}a_{ij}=-{10\over k}\;,
\label{ct4} \ee
where
$$
a^{+2} = D^{++} a^0 = {1\over 2}\;(D^{++})^2 a^{-2}\;.
$$
Note that the central charge $k$ is integer (if
we restrict ourselves to unitary representations of the $SU(2)$ Kac-Moody
algebra \cite{wit}), so eq. \p{ct4} means that $a^{ij}$ parametrize a
sphere $S^{2} \sim SU(2)/U(1)$, such that the reciprocal of its radius
is {\it quantized}.
Substituting this relation into eq. \p{r4}, where now $c^{1212} = {2\over 3}
a^{12}a^{12}$, one finds that it yields $a=4$, i.e. one of the three
integrable
cases of $N=2$ super KdV.

Dimension 3 charge is the hamiltonian itself.

In the case (B) one can construct a dimension 4 conserved charge.
It exists under the same restriction \p{ct4} on the square of $a^{ij}$
and reads:
\be H_4=\int [dZ]\;a^{-2}\;V^{++}(D^{--}V^{++})^2
+{i\over 6}\int [d\zeta^{-2}] \left[{7\over 6}\;(a^{-2})^3\;(V^{++})^4-k\;
a^{-2}\;( V^{++}_x)^2\right]. \label{ct5}
\ee

Thus we conclude that at least in the case (B) with the restriction \p{ct4}
our $N=4$ super KdV equation is expected to give rise to an integrable
hierarchy. Clearly, in order
to prove this one should either find the relevant Lax pair or prove the
existence of an infinite number of the conserved charges of the type
given above. We will address the problem of existence of the Lax
representation
for eq. \p{kdv4} in the nearest future.
In order to learn whether the other two integrable cases of $N=2$ super KdV
equation, with $a=-2$ and 1, possess integrable $N=4$ counterparts, one
should consider the dimension 5 conserved charge. In the $N=2$ case it
exists for all three integrable super KdV equation. A work on checking its
existence in the $N=4$ case is now in progress.

Finally, we wish to show that the $N=4$ super KdV equation with $c^{-4}
= (a^{-2})^2$ and the restriction \p{ct4} is bi-hamiltonian like its
$a=4$ $N=2$ prototype \cite{a7}. This is one more argument in favour of its
integrability.

The first hamiltonian structure of the $N=4$ super KdV
equation is associated with a Poisson bracket
which is obtained from the original one (\ref{poi}) by shifting
the superfield $V^{++}$ as follows
\be V^{++}\longrightarrow V^{++}+\beta a^{+2}(u)\;,\;\;\; \beta = const\;, \ee
and so it is proportional to the $SU(2)$
breaking parameter $a^{ij}$
\be
\left\{V^{++}(1), V^{++}(2)\right\}_{(1)}=\beta
\left( a^{0}(1) - a^{+2}(1){u^-_1u^+_2\over u^+_1u^+_2}
\right)
(D^+_1 \bar D^+_1)(D^+_2 \bar D^+_2)
\Delta(1-2) \label{poi1}\;.
\ee
Then, we take as a hamiltonian
the conserved charge $H_4$:
\be
H_{(1)}={9k\over 4\beta}\;H_4\;.\label{ha1}
\ee
The hamiltonian flow:
\be V^{++}_t=\{H_{(1)},V^{++}\}_{(1)} \label{kdv41} \ee
again yields the equation (\ref{kdv4}). Now this comes about in a rather
non-trivial way,
since both the new Poisson bracket (\ref{poi1})
and the new hamiltonian (\ref{ha1}) are proportional to the
$SU(2)$ breaking parameter $a^{ij}$, while the super
KdV equation (\ref{kdv4}) includes terms containing
no dependence on $a^{ij}$.
The key point is that these terms appear in (\ref{kdv41})
multiplied by the factor
\be -{k\over 10}\;s= -{k\over 20} \;a^{ij}a_{ij},
\ee
which is independent of harmonic coordinates $u^\pm$ and is
constrained to be $1$ from the requirement of
conservation of $H_2$ and $H_4$ (see eq. \p{ct4}).
\vspace{0.5cm}

{\bf 6. Conclusions.} To summarize, starting with $1D\;N=4$ harmonic superspace
we have defined the
supercurrent which generates $N=4\;SU(2)$ super Virasoro algebra via
an appropriate
Poisson structure \p{poi} and have constructed the
$N=4\;SU(2)$ super KdV equation \p{kdv4} as an
evolution equation for the supercurrent, with the $N=4\;SU(2)$ Poisson
structure as a second hamiltonian structure. We gave necessary
criteria for integrability of this equation and argued that it is
integrable at least for
one special choice of the $SU(2)$ breaking parameters in the hamiltonian,
such that these parameters describe a sphere $S^{2}$ with the radius subject
to the
quantization condition \p{ct4}.
This
particular $N=4$ super KdV equation
is bi-hamiltonian and,
upon the reduction $N=4 \Rightarrow N=2$,
yields the
$a=4$ integrable case
of the $N=2$ super KdV equation.
In order to prove its integrability and to learn whether two other integrable
$N=2$ super KdV equations (with $a=-2$ and 1) allow a generalization to
integrable
$N=4$ ones, it is of crucial necessity to construct the relevant Lax
representations and (or) to
inspect in more detail the issue of existence of higher order conserved
charges for eq.\p{kdv4} (beginning with $H_5$).

Besides these purely technical problems, the present study poses a number of
other interesting questions. For instance, one may wonder what are the $N=4$
analogs of Miura map and mKdV equation. This question could hopefully be
answered using free field representations of the $N=4\;SU(2)$ supercurrent,
e.g. of the kind given in ref. \cite{{a11},{a13}}. Another problem is to
re-derive eq.\p{kdv4} from the self-duality constraints of some
higher-dimensional supergauge theory along the lines of refs. \cite{a14}.

The harmonic superspace
approach suggests, in its own right, some new directions of extending the
results
presented here. An intriguing possibility is to discard the harmonic
constraint \p{ct3} for the supercurrent $V^{++}$, still keeping the
analyticity conditions \p{ct2}. Then
$V^{++}$ will be an unconstrained analytic $N=4$ superfield generating some
infinite dimensional higher isospin extension of $N=4\;SU(2)$ SCA. It is
interesting to inquire what are the relevant superfield Poisson bracket and
super KdV equation. Also, it seems that the harmonic superspace language
ideally suits for constructing the $W$ type nonlinear extensions of
$N=4\;SU(2)$ SCA and related generalized super KdV hierarchies.
\vspace{0.5cm}

{\bf Acknowledgements.} We sincerely thank Sergey Krivonos, Jean-Michel
Maillet and Oleg
Ogievetsky for interest in the work and useful discussions.

\end{document}